\documentclass[prl,aps,twocolumn,superscriptaddress]{revtex4-1}
\usepackage{amssymb,amsfonts,amsmath,graphicx,color,mathptm,times}
\usepackage{hyperref}
\usepackage{soul,cancel}
\newcommand{\ket}[1]{{\left\vert {#1} \right\rangle}}

\newcommand{\av}[1]{\langle#1\rangle}
\newcommand{\beq}{\begin{equation}}
\newcommand{\eeq}{\end{equation}}
\newcommand{\bea}{\begin{eqnarray}}
\newcommand{\eea}{\end{eqnarray}}

\newcommand{\tr}[1]{\textrm{Tr}\left[{#1}\right]} % Faster traces
\newcommand{\piu} {\!+\!}
\newcommand{\meno} {\!-\!}

\newcommand{\cf} {cf.~}
\newcommand{\ug} {\!=\!}
\newcommand{\eq}{Eq.~}
\newcommand{\eqs}{Eqs.~}
\begin{document}
\title{Landauer's principle in multipartite open quantum system dynamics}

\author{S. Lorenzo}
\affiliation{Dipartimento di Fisica e Chimica, Universit\`a degli Studi di Palermo, Via Archirafi 36, I-90123 Palermo, Italy}   
\author{R. McCloskey}
\affiliation{Centre for Theoretical Atomic, Molecular, and Optical Physics, School of Mathematics and Physics, Queen's University, Belfast BT7 1NN, United Kingdom}
\author{F. Ciccarello}
\affiliation{Dipartimento di Fisica e Chimica, Universit\`a degli Studi di Palermo, Via Archirafi 36, I-90123 Palermo, Italy}
\author{M. Paternostro}
\affiliation{Centre for Theoretical Atomic, Molecular, and Optical Physics, School of Mathematics and Physics, Queen's University, Belfast BT7 1NN, United Kingdom}
\author{G. M. Palma}
\affiliation{Dipartimento di Fisica e Chimica, Universit\`a degli Studi di Palermo, Via Archirafi 36, I-90123 Palermo, Italy}

\date{\today}

\begin{abstract}
We investigate the link between information and thermodynamics embodied by Landauer's principle in the open dynamics of a multipartite quantum system. Such irreversible dynamics is described in terms of a collisional model with a finite temperature reservoir. We demonstrate that Landauer's principle holds, for such a configuration, in a form that involves the flow of heat dissipated into the environment and the rate of change of the entropy of the system. Quite remarkably, such a principle for {\it heat and entropy power} can be explicitly linked to the rate of creation of correlations among the elements of the multipartite system and, in turn, the non-Markovian nature of their reduced evolution. Such features are illustrated in two exemplary cases.
\end{abstract}

\pacs{ insert pacs}

\maketitle

Logical irreversibility and heat dissipation are linked through the relation provided in 1961 by Landauer~\cite{Landauer}: the erasure of information on the state of a system is concomitant with the dissipation of heat into the surroundings. In turn, such heat is lower-bounded by a change in the information-theoretic entropy of the system. Landauer's principle, which has been recently tested~\cite{experiments}, was the building block of remarkable advances in our understanding of thermodynamics and its links with logical irreversibility and information theory~\cite{demone}. 

Recent progress in the quantum approach to non-equilibrium statistical mechanics~\cite{reviews,progress} has made the tracking of quantities such as heat, work, and entropy possible in an experimentally viable way. In turn this has enabled the test-bed demonstration of the link between information and energy in quantum systems subjected to elementary quantum computation protocols~\cite{recent}. The relation between information and the thermodynamic costs of quantum operations has been extensively addressed in the recent past. Noticeable examples include Refs.~\cite{Deffner}, (cf.~\cite{John} for a recent overview). Techniques of quantum statistical mechanics have been used to prove that a finite-size environment can provide tighter bounds to the heat generated in an erasure process~\cite{reeb,goold}. The validity of Landauer's principle has been verified for a quantum harmonic oscillator strongly coupled to a bath of bosonic modes~\cite{hilt}. However, such conceptual and experimental progress has not yet resulted in a satisfactory microscopic quantum framework able to account for the emergence of Landauer's principle. 

In this Letter we provide a derivation of a Landauer-like principle that addresses the heat and entropy fluxes implied in an erasure process described by a collision-based picture of open-system dynamics~\cite{CMs}. Such a microscopic formulation has been used to examine the process of thermalization of a quantum system in contact with a non-zero temperature bath and to investigate the link between with non-Markovianity and quantum correlations~\cite{Palma_2012, Ciccarello, McCloskey}. Here, we bring together these two perspectives: We analyse the process of information-to-energy conversion in collisional models that make use of general, non-restrictive assumptions and effective mechanisms to describe the heat dissipation into an environment. %that is deeply rooted into a microscopic picture. 
Our study thus aims at bridging the gap highlighted above and contributes to the ongoing research for a satisfactory foundation of Landauer's principle. 
The advantages of our approach are manyfold. First, modelling the irreversible dynamics of a system by a collisional model where, at each step, the environment is reset in a thermal state enables the formulation of Landauer's principle in terms of (heat and entropy) powers. Second, we show that the amount of heat dissipated by a multipartite quantum system is lower-bounded by the correlations established among the elements of the system itself. Finally, the flexibility of the picture adopted here allows us to unveil interesting links %with the memory-bearing nature of the dynamics experienced by part of such a system. We can thus connect the 
between the violation of Landauer's principle and the emergence of memory-bearing effects leading to non-Markovian dynamics.

\noindent
%--------------------------------------------------
{\it Erasure through thermalization.--}
%\label{single}
%--------------------------------------------------
We focus on the thermalization process of a system ${\cal S}$ in contact with an environment ${\cal R}$ endowed with a large number of degrees of freedom, which can thus be treated as a bath~\cite{CMs}. 
In our model $R$ consists of a collection of $N$ identical non interacting elements $\mathcal{R}_{~n}$, hence after dubbed  {\it subenvironments}, each of which is assumed to be in a thermal state. The bath is therefore in the product thermal state $\eta{=}\otimes^N_{n=1}\eta^{\rm th}_{n}$ with $\eta^{\rm th}_n=e^{{-}\beta \hat H_{\mathcal{R}_{~n}}}/{\rm Tr}[e^{{-}\beta \hat H_{\mathcal{R}_{~n}}}]$, $\hat H_{\mathcal{R}_{~n}}$ the free Hamiltonian of the $n^{\rm th}$ subenvironment, and $\beta$ the inverse temperature. We call $\hat H_{\mathcal{S}}$ the free Hamiltonian of the system and, with no loss of generality, we assume $\hat H_{\mathcal{R}_{~n}}=\hat H_{\mathcal{R}}$ for any $n$.
The system interacts with the environment via a sequence of pairwise collisions with individual subenvironments. The assumption of a big reservoir implies that the system never interacts twice with the same subenvironment, therefore at each collision the state of the subenvironment is refreshed and the system always interacts with subenvironments in a thermal state.  Each collision is described by a unitary operator $\hat U\!=\!e^{{-}i g \hat V \tau }$, where $g$ is a coupling constant,  $\tau$ is the collision time and  $\hat V{\!=\!}\sum_k \hat S_k{\otimes}\hat R_k$ is the (dimensionless) $\mathcal S$-$\mathcal {R}_{~n}$ interaction Hamiltonian [$\hat S_k$ ($\hat R_k$) are Hermitian operators acting on $\mathcal{S}$ ($\mathcal{R}_{~n}$)]. The information initially encoded in $\rho$ is gradually diluted into $\mathcal R$. 
After the $(n\piu1)^{\rm th}$ collisions, we find the states $\rho_{n+1}$ of the system and $\eta_{n+1}$ of the environment
\begin{eqnarray}
  \rho_{n+1}&=&\text{Tr}_{\cal R}[\hat U\,\rho_n\otimes\eta_n\,\hat U^{\dagger}]=\Phi[\rho_n],\label{Phi}\,\\
  \eta_{n+1}&=&\text{Tr}_{\cal S}\;[\hat U\,\rho_n\otimes\eta_n\,\hat U^{\dagger}]=\Lambda_{n}[\eta_n],\label{Lambda}
\end{eqnarray}
where $\Phi$ $(\Lambda_{n})$ is a completely positive trace-preserving (CPTP) map on $\mathcal S$ ($\mathcal R$)~\cite{Breuer02}. Unlike $\Phi$, map $\Lambda_{n}$ does depend on $n$ through $\rho_n$. 
% while either of them parametrically depends on $\hat U$ and $\eta$, 
A formal definition of mean heat flux and work is obtained considering infinitesimal changes of internal energy as $\delta {\rm Tr}[\mu\, \hat H]={\rm Tr}[\mu\, \delta\hat H]+{\rm tr}[\delta\mu\,\hat H]$, where $\mu$ is the density matrix of a system with Hamiltonian $\hat H$~\cite{varieU}. The first term is identified with the average work done on/by the system; the second, which disappears for a unitary evolution, is a heat flux. Such arguments have been more recently reprised in Refs.~\cite{varieU2}. In this sense, based on Eqs.~\eqref{Phi} and \eqref{Lambda}, the corresponding energy variation of the system $\Delta E_{n+1}$ and heat exchange with reservoir $\Delta Q_{n+1}$, are given by
\begin{equation}
\!\!\Delta E_{n{+}1}\!=\!{\rm Tr}\left[\hat H_{\cal S}(\Phi{-}\mathbb{I})[\rho_n]\right],\,\,\Delta Q_{n{+}1}\!=\!{\rm Tr}\left[\hat H_{\cal R}(\Lambda_{n}{-}\mathbb{I})[\eta]\right].\nonumber
\label{deltaEQ}\end{equation}
%We then expand the map $(\Phi-\mathbb{I})$ with respect to $y\!=\!g\tau$ as
%$\Phi-\mathbb{I}\!=\!\sum_{k=1}^\infty{y^j}\mathcal{K}_j/{j!}$
%with $\mathcal{K}_j={\partial^j_y\Phi}$ 
 We next assume $\av{\hat R_k}_{\eta_k}{\!=\!}{\rm Tr}_{\cal R}[\hat R_k\eta_k]{\!=\!}0$ (this is possible by moving into an interaction representation with respect to a suitably redefined local Hamiltonian of ${\cal S}$ and without affecting our results).  Assuming that each collision lasts a short time,  we can expand $(\Phi-\mathbb{I})$ in series up to second order in $y=g\tau$ to find $\Delta\rho_{n+1}=(\Phi-\mathbb{I})\rho_n=\!\mathcal K_2\,\rho_n$ with $\mathcal K_{\,2}={\rm Tr}_{\cal R}[\hat V(\rho_n\otimes\eta)\hat V-\{\hat V^2,\rho_n\otimes\eta\}/2]$. This gives us
\begin{eqnarray}
%\Delta\rho_{n+1}&=&(\Phi-\mathbb{I})\rho_n=\!\mathcal K_2\,\rho_n,\label{deltarhon}\\
	\Delta E_{n{+}1}&=&g^2\tau^2\sum_{kj}\av{\hat R_k \hat R_j}_\eta \langle\hat S_k \hat H_{\cal S} \hat S_j{-}\tfrac{1}{2}\{\hat S_k \hat S_j,\hat H_{\cal S}\}}\rangle_{\rho_n,\,\label{deltaEn}\\
   \Delta Q_{n{+}1}&=& g^2\tau^2\sum_{kj}\av{\hat S_k\hat S_j}_{\rho_{n}}\av{\hat R_k \hat H_{\cal R} \hat R_j{-}\tfrac{1}{2}\{\hat R_k \hat R_j,\hat H_R\}}_\eta.\label{deltaQn}
\end{eqnarray}
%where $\mathcal K_{\,2}={\partial^2_{y}\Phi}/{2}={\rm Tr}_{\cal R}[\hat V(\rho_n\otimes\eta)\hat V-\{\hat V^2,\rho_n\otimes\eta\}/2]$.
In the limit  $\tau\simeq0$ and $n\gg1$,  the quantity $t\!=\!n \tau$ becomes a continuous variable,  $\rho_n\!\to\!\rho(t)$ and $(\rho_{n+1}\!-\!\rho_n)/\tau\!\to\!\dot\rho$, so that we get the master equation (ME) $\dot{\rho}\!=\!\mathcal{K}_{\, 2}[\rho(t)]$. Likewise, Eqs.~(\ref{deltaEn}) and (\ref{deltaQn}) become
%\begin{equation}
%\dot{\rho}=\lim_{\substack{n \to \infty\\\tau \to 0}} \frac{\rho^{n{+}1}\!-\!\rho_n}{\tau}=\lim_{\substack{n \to \infty\\\tau \to 0}} \frac{(\Phi^\tau_\eta-\mathbb{I})\rho_n}{\tau}
%\end{equation}
\begin{eqnarray}
     &&\dot{E}=\gamma\sum_{k,j}\av{\hat R_k \hat R_j}_\eta\av{\hat S_k \hat H_{\cal S} \hat S_j{-}\tfrac{1}{2}\{\hat S_k \hat S_j,\hat H_S\}}_{\rho }\,,\\
     &&\dot{Q}=\gamma\sum_{k,j}\av{\hat S_k\hat S_j}_{\rho }\av{\hat R_k \hat H_{\cal R} \hat R_j{-}\tfrac{1}{2}\{\hat R_k \hat R_j,\hat H_R\}}_\eta\,,
\end{eqnarray}
where  we have defined the rate $\gamma\!=\!g^2\tau$, achieved by taking both $g\to\infty$ and $\tau\to0$ so that $g^2\tau$ is a constant. By doing so, we would guarantee the system-environment effects to persist (differently from what is obtained by taking $\tau\to0$). As discussed in Refs.~\cite{Palma_2012,Ciccarello}, with the assumption on $\av{\hat R_k}_{\eta}$ invoked above, such a continuous-time limit is  fully consistent and legitimate. By construction, $\rho(t)$ evolves according to a CPTP dynamical map~\cite{CMs,Breuer02}. 
%We now assume that $[\hat U^\tau,(\hat H_{\cal S}+\hat H_{\cal R})]{=}0$, so that $\dot{Q}(t)=-\dot{E}_{\cal S}(t)$.
We now focus on energy-conserving $\mathcal S$-$\mathcal R$ interactions and assume $[\hat U,(\hat H_{\cal S}\!+\!\hat H_{\cal R})]{=}0$. Hence, $\dot{Q}\!=\!-\dot{E}$ and the stationary state of the system is the Gibbs state at the same initial temperature of the bath, {\it i.e.} $\rho^{eq}{=}e^{-\beta H_{\cal S}}/\tr{e^{-\beta H_{\cal S}}}$. In this case, the stationary state is such that $\dot{E}=\dot{Q}=0$, which is not verified by non-energy-conserving models. Is such cases, the steady state corresponds to a constant flux of heat into/out of the system.

The relative entropy $S(\rho|\rho^{eq})$ between the state at time $t$ and the stationary one obeys the relation
$\dot S(\rho|\rho^{eq})=%\tr{\dot{\rho} (\ln \rho-\ln \rho^{eq})}=
-\dot{S}(\rho)+\beta \dot{E}$~\cite{Rosario2013}.
Due to the non-increasing nature of the relative entropy under CPT maps~\cite{Vedral2002}, 
we obtain
\begin{equation}
\beta \dot{Q}(t)\geq\dot{\tilde{S}}(\rho),
\label{bound}
\end{equation}
where $\tilde{S}(\rho)=-S(\rho)$. This embodies an open-system formulation of Landauer's principle for the heat and entropy {\it fluxes}. We can thus  lower-bound the flow of heat dissipated into the environment with the rate of change of a key information theoretical quantity. Remarkably, our formulation enables the assessment of the {nature} of the open-system dynamics of ${\cal S}$. The power of our formalism is thus twofold: on one hand, it allows for a time-resolved analysis of the erasure-by-thermalization process. On the other hand, it elucidates the role of correlations in the efficiency of information-erasure processes. This is illustrated by two examples of irreversible collisional dynamics. We first analyse a cascaded system where the subenvironments interact in sequence with the elements of a bipartite system, chosen here for easiness of presentation. In this scenario, the environment is ``recycled" to erase the information in the two subsystems before being damped. We show that the efficiency of information erasure is limited by the build-up of intersystem correlations. In the second example, which we dub {\it indirect}, we  consider a system interacting with an ancilla whose state is then erased by the environment. In this case, we show the appearance of local violation of Landauer's Principle due to the onset of memory effects.

%--------------------------------------------------
\noindent
{\it Information erasure in cascaded systems.--}
%\label{recycle}
%--------------------------------------------------
For now, we consider the system $\mathcal S$ as {bipartite} and consisting of two identical subsystems ${\cal S}_{A,B}$. Each subenvironment $\mathcal R_{~n}$ collides with ${\cal S}_A$ first and ${\cal S}_B$ later. Each collision is modelled as a unitary process with associated evolution operator $\hat U_{X}{=}e^{-i g \hat V_{X}\tau}$, ($X{=}A,B$) and generator $\hat V_{X}{=}\sum_{k} \hat S_{Xk}{\otimes}\hat R_{k}$ (here $\hat S_{Xk}$ acts on subsystem $X$).
%As before, ${\mathcal R}$ is initially in state $\eta$, while ${\cal S}$ is initialised in a generic $\rho^0_{AB}$. 
A ``step" consists of an $\mathcal S_A$-$\mathcal R_{\,n}$ collision followed by an $\mathcal S_B$-$\mathcal R_{~n}$ one. Hence, the joint $\mathcal S$-$\mathcal R$ state at step $n\!+\!1$ is given by 
$\sigma_{n+1}{=}\hat U_B \hat U_A \rho_{n}\eta\hat U_A^{\dagger}\hat U_B^\dag$ ($\rho_n$ is the joint ${\cal S}_A$-${\cal S}_B$ state). \eqs(\ref{Phi}) and (\ref{Lambda}) thus still hold with $\hat U\ug \hat U_B\hat U_A$.

\begin{figure*}[ht!]
\begin{center}
\hskip0cm{\bf (a)}\hskip4cm{\bf (b)}\hskip4cm{\bf (c)}\hskip4cm{\bf (d)}
\includegraphics[width=2.0\columnwidth]{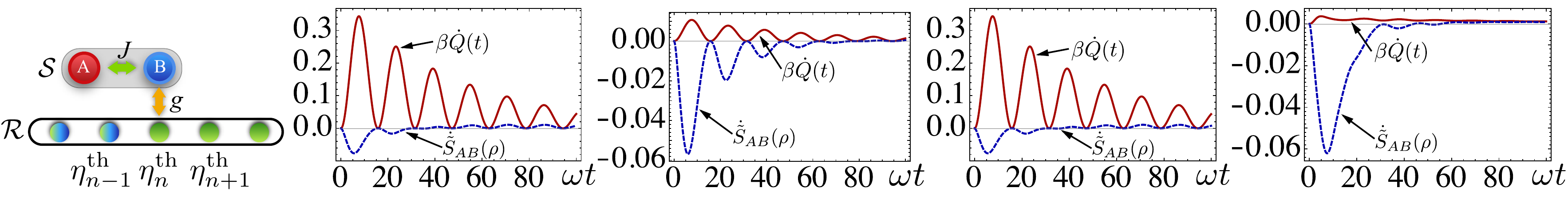}
\caption{(Color online) {\bf (a)} Sketch of the recycled environment setup: after colliding with ${\cal S}_A$, subenvironment ${\cal R}_{~n}$ (prepared in  $\eta^{\rm th}_n$) is used to erase the state of ${\cal S}_B$. %At step $n$, subenvironment ${\cal R}_{~n}$ is prepared in $\eta_n$ and interacts sequentially with ${\cal S}_A$ and ${\cal S}_B$ through the unitaries $\hat U_{X=A,B}$. 
By tracing out the subenvironments, a Markovian ME is achieved for ${\cal S}$. {\bf (b)} Heat flux and entropy for a system composed of a single qubit for $\xi\ug0.9$ and $\xi=-0.9$ [inset]. {\bf (c)} Heat and entropy flux for the cascade model. We have set $\gamma/\omega\ug1$ with ${\cal S}$ initially prepared in $\ket{\uparrow\uparrow}$ at $\xi\ug0.9$ and $\xi=-0.9$ [inset]. {\bf (d)} Mutual information $I(A:B)$ between the subsystems in the same case of panels {\bf (b)} and {\bf (c)} for $\xi{=}0.9$ (solid brown line) and $\xi{=-}0.9$ (dashed green line). Heat fluxes are expressed in unit of $\omega\gamma$.}
\label{default}
\end{center}
\end{figure*}

Through a procedure analogous to the one outlined earlier, it can be shown that in the continuous-time limit the bipartite system ${\cal S}$ is governed by the ME~\cite{Palma_2012}
\begin{equation}
\dot{\rho }=
\sum_{X=A,B}\mathcal{L}_X[\rho ]+\mathcal{D}_{AB}[\rho ],\label{ME}
\end{equation}
where superoperators $\mathcal{L}_X$ and $\mathcal{D}_{AB}$ are defined by
\begin{eqnarray}
\mathcal{L}_X[\rho ]&\ug&\gamma\sum_{kj}\av{\hat R_k\hat R_{j}}_{\eta}(\hat S_{Xj}\rho  \hat S_{Xk}{-}\tfrac{1}{2}\lbrace \hat S_{Xk}\hat S_{Xj},\rho \rbrace),\label{LX}\\
\mathcal{D}_{AB}[\rho ]&\ug&\gamma\sum_{kj}\av{\hat R_j\hat R_k}_{\eta}\left[\hat S_{Ak}\rho ,\hat S_{Bj}\right]{+}
			               \av{\hat R_k\hat R_j}_{\eta}\left[\hat S_{Bj},\rho  \hat S_{Ak}\right].\,\,\,\,\,\,\,\,\,\,\,\label{DAB}
\end{eqnarray}
Subsystems $\mathcal S_A$ and $\mathcal S_B$ {\it jointly} undergo a Markovian dynamics. %, where $\mathcal K_{~2}\ug \sum_x \mathcal L_{X}\piu \mathcal D_{AB}$. 
This is true also for the dynamics of $\mathcal S_A$, given that it collides  with ``fresh" subenvironments every time~\cite{farace}. However, in general, $\mathcal S_B$ is subjected to non-Markovian evolution. 

As before, given that no work is performed on the system, we identify the variation rate of the internal energy of the system $\dot{E}$ with 
$\Delta E_{n{+}1}\!=\!{\rm Tr}\left[\hat H_{\cal S}(\Phi{-}\mathbb{I})[\rho_n]\right]$. 
Using \eq(\ref{ME})
%this reads
%\begin{equation}
%\begin{aligned}
%\dot{E}
%\ug\rm{Tr}\!\!\left[\left(\!\sum_{X=A,B}\mathcal{L}_X[\rho ]\piu\mathcal{D}_{AB}[\rho ]\right)\!\!\sum_{X=A,B}\hat H_{{\cal S}_X}\right]\\
%%&=\sum_{X=A,B}\dot{E}_{X}(t)+\dot{E}_{AB}(t)
%\end{aligned}
%\end{equation}
%with $\hat H_{{\cal S}_X}$ the free Hamiltonian of subsystem $X=A,B$.
and following an approach similar to the one presented above, it is possible to write explicitly the expression for the heat transferred at each step of the collision-based process, 
as shown in the Appendix.
%This is given by the quantity  $\Delta Q_n{=}{\rm Tr}[\hat H_{\cal R}(\eta^n-\eta)]$.
%When taking the continuous limit of $n\to\infty$ and $\tau\to0$, we find that the rate of change of the heat exchanged with the environment reads
%\begin{equation}
%\begin{aligned}
%  \dot{E}&=\gamma\Big[\sum_{kj}\alpha^R_{kj}(\av{\hat S_{Ak}\hat S_{Aj}}_{\rho }+\av{\hat S_{Bk}\hat S_{Bj}}_{\rho })\\
%  &+2\text{Re}(\av{\hat S_{Ak}\hat S_{Bj}}_{\rho }\av{\hat R_j[\hat H_{\cal R},\hat R_k]}_\eta)\Big].\\
%\end{aligned}
%\end{equation}
By focusing again on energy-conserving system-environment interactions, the identity $\dot{Q}\ug-\dot{E}$ holds and the system thermalizes at the same temperature of the environment. Accordingly, in light of the absence of any direct $\mathcal S_{A}$-$\mathcal S_{B}$ coupling, the system asymptotically reaches the stationary state $\rho^{eq}\ug\rho_A^{eq}\otimes\rho_B^{eq}={e^{-\beta H_{{\cal S}_A}}}\!\otimes\!{e^{-\beta H_{{\cal S}_B}}}/{(Z_AZ_B)}$ (here $Z_X={\rm Tr}[e^{-\beta\hat H_{{\cal S}_X}}]$), i.e. the a factorised state of locally thermal states at the same temperature. Indeed, despite the $\mathcal S_{B}$ evolution depends on $\mathcal S_{A}$ (thus determining its non-Markovian character), $\mathcal S_{A}$ thermalizes with $\mathcal R$ in a fully Markovian fashion. Thereby, asymptotically, the $\mathcal S_{A}$-$\mathcal R_{\,n}$ collisions do not change the $\mathcal R_{\,n}$ state. Thus, in the same limit, $\mathcal S_{B}$ will also collide with subenvironments that are still in state $\eta$, forcing it to thermalize at the temperature of the environment.
We remark that the bound stated in Eq.~(\ref{bound}) clearly holds in the present situation, provided that $Q$ and $\tilde S(\rho)$ refer to system $\mathcal S$.
%Using $\dot S(\rho|\rho^{eq})=
%-\dot{S}(\rho)+\beta \dot E^{tot}_{AB}$~\cite{Rosario2013},
%we finally find
%\bea
%\beta \dot{Q}(t)\geq-\dot{S}_{AB}(\rho) =\dot{\tilde{S}}_{AB}(\rho).
%\label{Land1}
%\eea

In order to show that recycling the environment % - i.e. reusing a subenvironment after it has collided with ${\mathcal S_A}$ - 
affects the erasure efficiency, we use the definition of quantum mutual information 
${\cal{I}}(\mathcal S_A:\mathcal S_B)=S(\rho_{\mathcal S_A}){+}S(\rho_{\mathcal S_B})-S(\rho)$ 
and conditional quantum entropy $S_{\mathcal S_A|\mathcal S_B}\ug S(\rho){\!-\!}S(\rho_{\mathcal S_A})$ \cite{nc}, to get 
$-\dot{S}(\rho) =-2\dot{S}(\rho_{\mathcal S_A})+\dot{S}_{\mathcal S_A|\mathcal S_B}\meno\dot{S}_{\mathcal S_B|\mathcal S_A}\piu \dot{\cal{I}}(\mathcal S_A{:}\mathcal S_B)$ and, in turn %(we recall that $\tilde{S}\ug-S$)
\bea
\beta \dot{Q}>2\dot{\tilde{S}}(\rho_{\mathcal S_A})+\dot{S}_{\mathcal S_A|\mathcal S_B}-\dot{S}_{\mathcal S_B|\mathcal S_A}+\dot{I}(\mathcal S_A{:}\mathcal S_B).
\label{Land2}
\eea
This shows that the erasure efficiency in cascaded systems depends on the rate at which correlations between $\mathcal S_A$ and $\mathcal S_B$  are established. Indeed, all the quantities in Eq.~\eqref{Land2} are null only for product states. 
Incidentally, the rate $2\dot{\tilde{S}}(\rho_{\mathcal S_A})$ is that of two identical state erased by independent reservoirs.
\begin{figure*}[t]
\begin{center}
{\bf (a)}\hskip3cm{\bf (b)}\hskip3cm{\bf (c)}\hskip3cm{\bf (d)}\hskip3cm{\bf (e)}
\includegraphics[width=2.0\columnwidth]{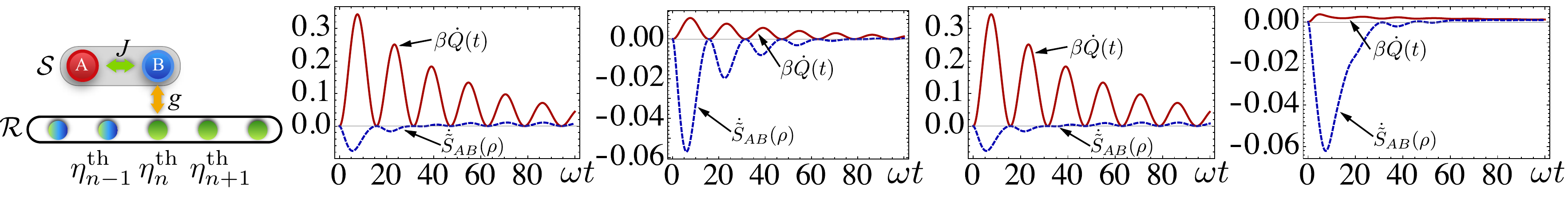}
\caption{(Color online) {\bf (a)} Sketch of the indirect erasure protocol: the bipartite system ${\cal S}$ consists of subsystems ${\cal S}_A$ and ${\cal S}_B$. After ${\cal S}_B$ interacts with the $n^{\rm th}$ subenvironment (prepared in $\eta^{\rm th}_n$), it collides with ${\cal S}_A$ and is then directed to element ${\cal R}_{~n+1}$. ${\cal S}_B$ thus bridges the erasure process undergone by ${\cal S}_A$, which experiences a non-Markovian evolution. {\bf (b)-(e)}: Total heat flow $\beta \dot Q$ against the entropy change rate $\dot{\tilde S}$ for a qubit-qubit ({\bf b},{\bf c}) and qubit-harmonic oscillator ({\bf d},{\bf e}) configurations. We have set $J/\omega\ug0.1$ and $\gamma_g/\omega\ug0.01$, taking $\mathcal{S}_B$ at the same temperature of the environment, $\beta{=}10$ ({\bf b},{\bf d}) and $\beta{=}0.5$ ({\bf c},{\bf e}). A study against the value of $J$ and $\gamma_g$ is reported in the Appendix.
}
\label{QQheatflow}
\end{center}
\end{figure*}

Our arguments are straightforwardly extended to the multipartite case. Using the correlation information $\mathcal{I}_N=\sum_{i=1}^N S_i \meno S_{1...N}$~\cite{Herburt2004}, which extends mutual information to an $N-$partite system, Eq.~\eqref{Land2} is extended to the erasure of $N$ copies as
\bea
\beta \dot Q>N\dot{\tilde{S}}(\rho_1)\piu(N\meno1)\dot{S}_{1|2...N}\meno\textstyle\sum_{k=2}^N \dot{S}_{k|1...}\piu\dot{\mathcal{I}}_N\,.
\label{LandN}
\eea
We now illustrate the results described until now with a specific example. We consider an all-qubit configuration where $\hat H_{\cal S}{\ug}(\omega/2)\sum_{X=A,B}\hat \sigma_{Xz}$ and $\hat H_{{\cal R}_{~n}}\ug(\omega/2)\hat \sigma_{{\cal R}_{~n}z}$ with $\{\hat \sigma_{\ell i }\}$ ($i\ug x,y,z$) the Pauli operators of qubit $\ell\ug X,\mathcal{R}_{\,n}$.
Each subenvironment is prepared in the thermal state
$\eta{\ug}[({1\meno\xi})/{2}]|{\uparrow}\rangle\langle {\uparrow}|\piu[({1\piu\xi})/{2}]|{\downarrow}\rangle\langle {\downarrow}|$, where $\{\ket{{\uparrow}},\ket{{\downarrow}}\}$ are the eigenstates of $\hat\sigma_z$ with $\xi{=}\tanh({{\beta\omega}/{2}})$.
The ${\cal S}_X$-$\mathcal R_{\,n}$ interaction Hamiltonian is of the isotropic $XX$ form, i.e., {$\hat V_X{\ug}\sum_{i=x,y}\hat \sigma_{Xi}\otimes\sigma_{\mathcal{R}_{\,n}i}$}. 
In the Appendix we present the form taken by Eq.~\eqref{ME} in this case and show that the evolution of ${\cal S}_A$ is Markovian and not causally related to ${\cal S}_B$. 
Instead, the $\mathcal S_B$  dynamics depend crucially on the state of $\mathcal S_A$, thus being steered by previous subsystem-environment collisions. 
%The heat flow from (or into) ${\cal S}$ can be conveniently expressed \cite{supp} in terms of the entries $\rho^{ij}_{kl}=\langle i,j|\rho |k,l\rangle$ ($i,j,k,l=\uparrow,\downarrow$) of the bipartite density matrix $\rho$. This reads
%\begin{equation}
%\dot Q=4\gamma\omega\left[ \xi\left(1 \piu 2{\rm Re}\left[ {\rho^{\uparrow\downarrow}_{\downarrow\uparrow}}\right]\right)\piu \left( {\rho ^{\uparrow\uparrow}_{\uparrow\uparrow}}\meno   {\rho^{\downarrow\downarrow}_{\downarrow\downarrow}}\right)\right],
%\end{equation}
%showing the direct dependence of the heat flow on the coherences of the $\mathcal S$ state. 
The behaviour of $\beta\dot Q$ is compared to $\dot{\tilde S}(\rho )$ in Fig.~\ref{default} for two different temperatures and each subsystem initially prepared in $\ket{\uparrow}$. To pinpoint the role played by the $\mathcal S_A$-$\mathcal S_B$ correlations in setting the non-monotonic temporal trend of the heat flow [cf. Fig.~\ref{default} {\bf (c)}], we have studied $I(A:B)$. Fig.~\ref{default} {\bf (d)} shows that the establishment of such correlations indeed acts as a precursor of the non-monotonicity of the heat exchanged with the environment. In the Appendix, we support these conclusions with the study of a multipartite system.

\noindent
%---------------------------------------------------
{\it Indirect Erasure.--}
\label{inderasure}
%---------------------------------------------------
We now analyse the dynamics of a bipartite system ${\cal S}$, in which the evolution of the two subsystems and their coupling with the subenvironments are treated on different footings: we include a mutual interaction between ${\cal S}_A$ and ${\cal S}_B$ while  we assume that only ${\cal S}_B$ interacts with $\mathcal R$. This amounts to assume that the information contained in  ${\cal S}_A$  is first transferred to   ${\cal S}_B$  and then damped into   $\mathcal R$: a clear example of indirect information erasure. In practice we assume that between the ${\cal S}_B$-$\mathcal{R}_{~n}$ and ${\cal S}_B$-$\mathcal{R}_{~n+1}$ collisions, the two subsystems ${\cal S}_A$ and ${\cal S}_B$ interact. \eqs(\ref{Phi}) and (\ref{Lambda}) are still valid provided that $\hat U\!\to\! \hat U_B\hat U_{AB}$, where $\hat U_B$ (describing an $\mathcal S_B$-$\mathcal R_{\,n}$ collision) is the same as in the last Section, while $\hat U_{AB}\ug e^{-i \hat{H}_{AB}\tau}$ regulates the joint evolution of the ${\cal S}_A$-${\cal S}_B$ system.
%\begin{equation}
%\label{indirect}
%\begin{aligned}
%\rho_{n+1} &= 
%{\rm Tr}_\mathcal{R}[\hat U^\tau_{BR}\hat U^\tau_{AB}(\rho_{n}\otimes\eta)\hat U_{AB}^{\tau\dag} \hat U_{BR}^{\tau\dag}]=
%\phi_\eta[\rho_n],\\ 
%\eta_{n+1}&=
%{\rm Tr}_\mathcal{S}[\hat U^\tau_{BR}\hat U^\tau_{AB}(\rho_n\otimes\eta)\hat U_{AB}^{\tau\dag} \hat U_{BR}^{\tau\dag}]=
%\lambda_{\rho_n}[\eta]
%\end{aligned}\eta_{n+1} 
%\end{equation}
%with $\phi$ a map that evolves the state of the bipartite system ${\cal S}$ and $\lambda(\rho_n)$ the effective dynamical process, for each subenvironment, 
%that changes at every step depending on the system's state. I
To go to continuous-time, we expand each unitary operator to lowest-order in $\tau$.
Unlike the previous case, we should now take %$\hat U_{AB}$ up to {\it first} order in $\tau$. In other words, here we approximate as 
$\hat U_{B}\simeq\mathbb{I}{-}i g\tau \hat V_{B}{-}\frac{g^2\tau^2}{2}\hat V^2_{B}$, as before, and $\hat U_{AB}\simeq\mathbb{I}{-}i \tau\, \hat H_{AB}$. We thus find% Replacing these into the analogous equation of \eq(\ref{deltarhon}) and assuming again $\av{\hat R_k}_{\eta}{=}0$, we find 
\begin{equation}
\label{mainmaster}
\dot\rho {=}
	-i[J\hat{H}_{AB},\rho ]
	{+}\gamma\sum_{kj }\av{\hat R_{k}\hat R_{j }}_\eta(\hat S_{Bj }\rho \hat S_{Bk}
	{-}\tfrac{1}{2}\lbrace \hat S_{Bk}\hat S_{Bj },\rho \rbrace).\!\!
\end{equation}
This is a Lindblad ME describing Markovian dynamics of $\mathcal S$. The dynamics of $\mathcal S_A$, however, is in general non-Markovian. 

\begin{figure}[b]
\begin{center}
{\bf (a)}\hskip4cm{\bf (b)}
\includegraphics[width=1.0\columnwidth]{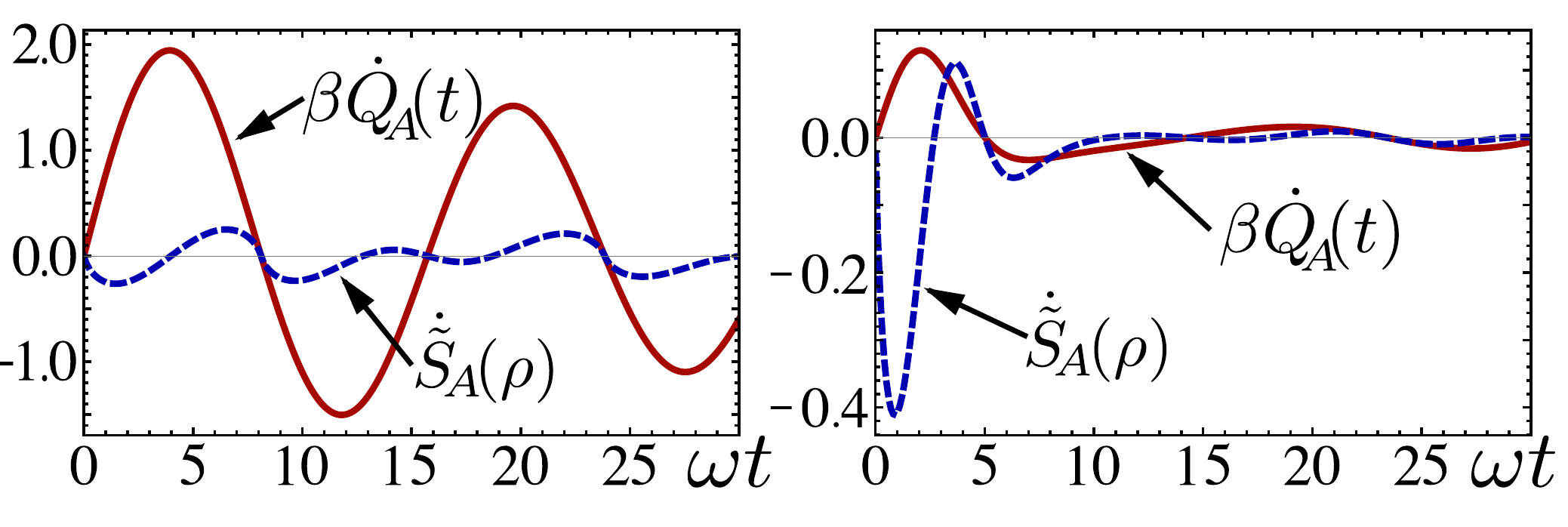}
\caption{(Color online) We plot $\beta\dot{Q}_A(t)$ and $\dot{\tilde S}_A$ for the qubit-oscillator configuration in the indirect-erasure case. In panel {\bf (a)} [{\bf (b)}] we have used $\beta=10$ [$\beta=0.5$] and the parameters stated in Fig.~\ref{QQheatflow} {\bf (d)} and {\bf (e)}.}
\label{HOQ}
\end{center}
\end{figure}

We now illustrate the indirect erasure model with a specific example where we set no constraint on the dimension of the Hilbert space of ${\cal S}_B$. In the Appendix, we derive the form of Eq.~\eqref{mainmaster} when ${\cal S}_B$ is a quantum harmonic oscillator resonantly coupled to a two-level subsystem ${\cal S}_A$ via the interaction Hamiltonian $\hat H_{AB}{=}\omega(\hat q^2{+}\hat p^2{+}\hat \sigma_{Az}/2){+}J(\hat q\hat \sigma_{Ax}{+}\hat p\hat \sigma_{Ay})$, with $\hat q$ and $\hat p$ the position and momentum quadrature operators of oscillator $\mathcal S_B$, respectively. An analogous model holds for the ${\cal S}_B-{\cal R}_{~n}$ interaction, each subenvironment being a qubit. The bosonic nature of $\mathcal S_B$ allows us to explore the effects that multiple excitations in ${\cal S}_B$ have on both the heat flow and rate of entropy change. 
Indeed, as illustrated in Figs.~\ref{QQheatflow} {\bf(d)} and {\bf(e)}, for an initially thermal harmonic oscillator at the same temperature of the bath, the differences with a finite-dimensional subsystem ${\cal S}_B$ are striking, and dependent on $\beta$.
At low temperature, due to the excitation-conserving nature of the Hamiltonian, 
the dynamics of ${\cal S}$ is in fact identical to that of a two-qubit system, as only one excitation at most can be exchanged with ${\cal R}$. 
This is not the case when the temperature is raised, as excitations can be pumped {\it into} the harmonic oscillator at each collision [\cf Fig.~\ref{QQheatflow} {\bf(e)}]. %, which is for $\beta=0.5$]. 
In the Appendix we study the behavior of the key quantities in our analysis against $J$ and $\gamma_g$. %The nature of the dynamics responsible for the indirect erasure process allows us to make an 
Interesting considerations on the nature of the dynamics of ${\cal S}_A$ are in order. Indeed, by tracing out the degrees of freedom of ${\cal S}_B$, we obtain a non-Markovian evolution due to the flowback of information into ${\cal S}_A$.  Correspondingly, Eq.~(\ref{bound}) is violated very quickly, as shown in Fig.~\ref{HOQ}, which again highlights the effects that the establishment of intra-system correlations have on the efficiency of the information-to-energy conversion processes. Similar conclusions hold when relaxing the key assumption of non-interacting subenvironments. The break-down of such condition would entail the establishment of memory effects within the environment and the occurrence of non-Markovianity in the ${\cal S}_A$-${\cal S}_B$ dynamics. The violation of our formulation of Landauer's principle could thus be used to infer the nature of the bath itself.

\noindent
{\it Conclusions.} We have investigated the information-to-energy conversion processes in terms of collision-based models to  describe the system-environment interaction, with the aim to go towards the a microscopic formulation of Landauer's principle. In our approach we focused on the rate of entropy change and dissipated-heat flux involved in the overall erasure process. By doing so, we have been able to link the efficiency of the erasure process to the intra-system correlations arising in the open dynamics of a multipartite system. 

\noindent
{\it Acknowledgments.} We thank T. J. G. Apollaro, A. Farace, O. Dahlsten, E. Lutz, and Y. Omar for discussions and suggestions. MP is grateful to J. Goold and K. Modi for fruitful discussions. We acknowledge financial support from the Department of Employment and Learning, the John Templeton Foundation (grant ID 43467), MIUR - PRIN2010/11 and the EU Collaborative Project TherMiQ (Grant Agreement 618074). This work was partially supported by the COST Action MP1209.

\section*{Appendix}

%--------------------------------------------------
%\section{Introduction}
%\label{intro}
%--------------------------------------------------
\subsection{Heat flux in the qubit `recycling environment' model}
As outlined in the main text,  no work is performed on the system, so we identify the variation rate of the $\mathcal S$ internal energy $\dot{E}$ with 
\begin{equation}
\!\!\Delta E_{n{+}1}\!=\lim_{\substack{n \to \infty\\\tau \to 0}}\frac{\!{\rm Tr}\left[\hat H_{\cal S}(\Phi{-}\mathbb{I})[\rho_n]\right]}{\tau},
%\,\,\Delta Q_{n{+}1}\!=\!{\rm Tr}\left[\hat H_{\cal R}(\Lambda_{n}{-}\mathbb{I})[\eta]\right].\nonumber
\label{deltaEQ}\end{equation}
Remembering the master equation 
\begin{eqnarray}
\dot{\rho }&\ug&\sum_{X=A,B}\mathcal{L}_X[\rho ]+\mathcal{D}_{AB}[\rho ],\nonumber\\
\mathcal{L}_X[\rho ]&\ug&\gamma\sum_{kj}\av{\hat R_k\hat R_{j}}_{\eta}(\hat S_{Xj}\rho  \hat S_{Xk}{-}\tfrac{1}{2}\lbrace \hat S_{Xk}\hat S_{Xj},\rho \rbrace),\nonumber\\
\mathcal{D}_{AB}[\rho ]&\ug&\gamma\sum_{kj}\av{\hat R_j\hat R_k}_{\eta}\left[\hat S_{Ak}\rho ,\hat S_{Bj}\right]{+}
			               \av{\hat R_k\hat R_j}_{\eta}\left[\hat S_{Bj},\rho  \hat S_{Ak}\right].\,\,\,\,\,\,\,\,\,\,\,\label{ME}
\end{eqnarray}
this reads
\begin{equation}
\begin{aligned}
\dot{E}
\ug\rm{Tr}\!\!\left[\left(\!\sum_{X=A,B}\mathcal{L}_X[\rho ]\piu\mathcal{D}_{AB}[\rho ]\right)\!\!\sum_{X=A,B}\hat H_{{\cal S}_X}\right]\\
\end{aligned}
\end{equation}
with $\hat H_{{\cal S}_X}$ the free Hamiltonian of subsystem $X{=}A,B$.
Following the recipe presented in the main text, it is possible to write explicitly the expression for the heat transferred at each step of the collision-based process, 
this is given by the quantity  $\Delta Q_n{=}{\rm Tr}[\hat H_{\cal R}(\eta_n-\eta)]$.
When taking the continuous limit of $n\to\infty$ and $\tau\to0$, we find that the rate of change of the heat exchanged with the environment reads
\begin{equation}
\begin{aligned}
  \dot{E}&=\gamma\Big[\sum_{kj}\alpha_{kj}(\av{\hat S_{Ak}\hat S_{Aj}}_{\rho }+\av{\hat S_{Bk}\hat S_{Bj}}_{\rho })
  +2\text{Re}(\beta_{kj}\av{\hat S_{Ak}\hat S_{Bj}}_{\rho })\Big].\nonumber
\end{aligned}
\end{equation}
%\begin{equation}
%\begin{aligned}
%  \dot{E}&=\gamma\Big[\sum_{kj}\alpha^R_{kj}(\av{\hat S_{Ak}\hat S_{Aj}}_{\rho }+\av{\hat S_{Bk}\hat S_{Bj}}_{\rho })\\
%  &+2\text{Re}(\av{\hat S_{Ak}\hat S_{Bj}}_{\rho }\av{\hat R_j[\hat H_{\cal R},\hat R_k]}_\eta)\Big].\\
%\end{aligned}
%\end{equation}
with $\alpha_{kj}{=}\av{R_k H_R R_j{-}\frac{1}{2}\{R_k R_j,H_R\}}_\eta$ and $\beta_{kj}{=}\av{\hat R_j[\hat H_{\cal R},\hat R_k]}_\eta$.
%--------------------------------------------------------------------------------------------------------------------------------------------------------------------------------------------------
\subsection{Example I: Master equation for the qubit `recycling environment' model}
\begin{figure}[t]
\begin{center}
%{\bf (a)}\hskip5.5cm{\bf (b)}\hskip5.5cm{\bf (c)}
\includegraphics[width=1.\columnwidth]{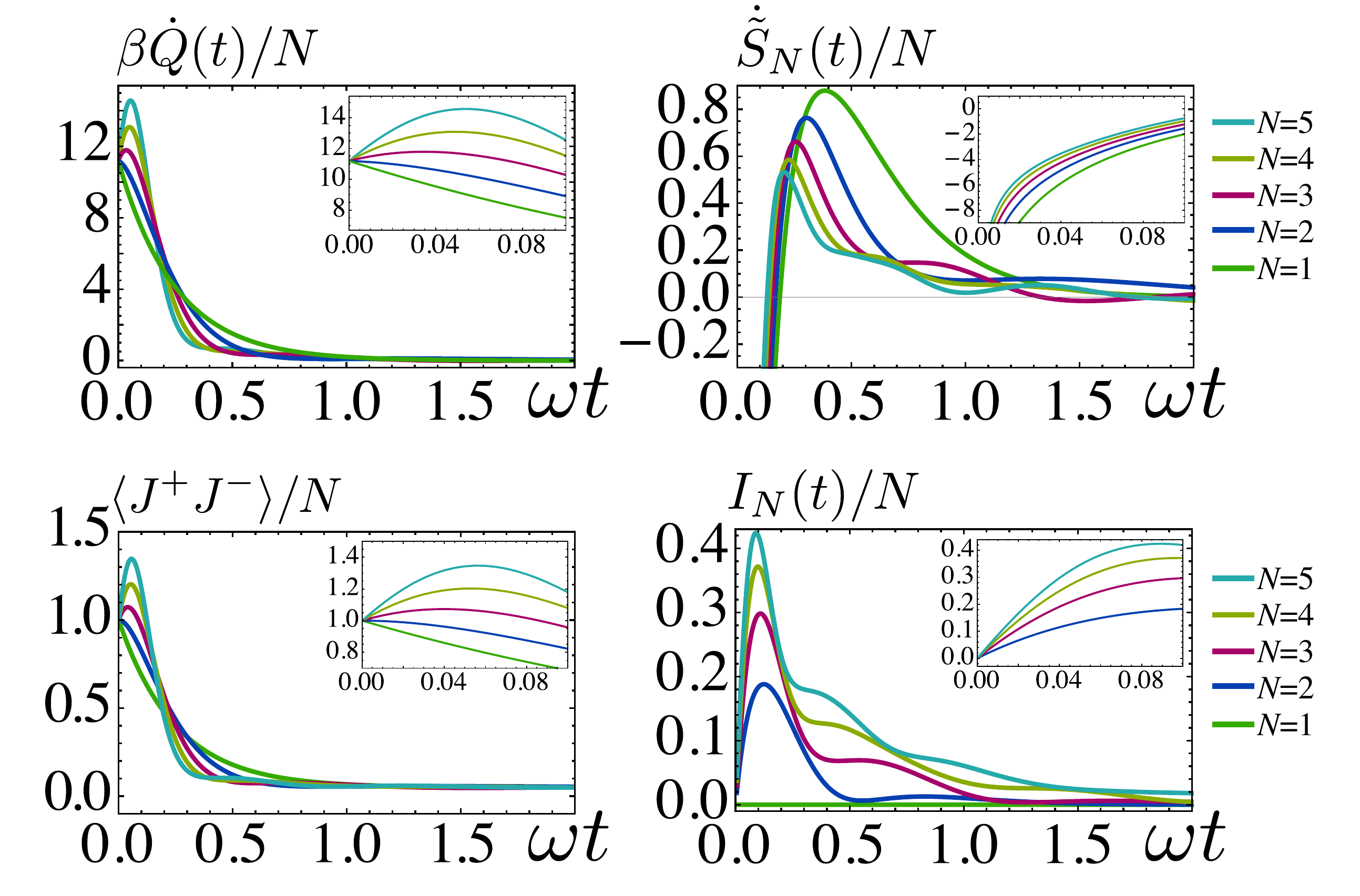}
\caption{(Color online) 
Recycling environment configuration for different numbers of subsystems $N$. We have used $\gamma/\omega=1$ with $\xi=0.9$. All the subsystems ${\cal S}_n$ are prepared in $\ket{\uparrow}$. All the quantities, i.e. heat flux $\beta\dot{Q}$, entropy variation $\dot{\tilde{S}}$, mutual information $I_N$ and radiation rate $\langle J^+J^-\rangle$, have been rescaled with $N$ for better visibility}
\label{supp_1}
\end{center}
\end{figure}

%--------------------------------------------------------------------------------------------------------------------------------------------------------------------------------------------------
As described in the main text each collision is generated by a Hamiltonian of the form $\hat H_{X}=\sum_{k} \hat S_{k}^{X}{\otimes}\hat R_{k}$. 
Here, $\hat S^X=\lbrace\sigma^X_x,\sigma^X_y\rbrace$ and $\hat R=\lbrace\sigma^\mathcal{R}_x,\sigma^\mathcal{R}_y\rbrace$.
We assume the sub environment $\mathcal{R}_{\;n}$ initially in a thermal state at temperature $\beta$, defining $\xi{=}\tanh(\beta\omega/2)$, this means 
$\eta^{\rm th}_n{=} \frac{e^{-\beta\frac{\omega}{2}\sigma^{\mathcal{R}_n}_z}}{\text{Tr}[e^{-\beta\frac{\omega}{2}\sigma^{\mathcal{R}_n}_z}]}{=}\left(\begin{smallmatrix}\frac{1-\xi}{2}&0\\0&\frac{1+\xi}{2}\end{smallmatrix}\right)$. 
Introducing the Pauli ladder operators $\hat \sigma^\pm_{q}{=}(\hat \sigma^x_{q}{\pm} i\hat \sigma^y_{q})/2$ for $q{=}{\cal S}_{A,B},{\cal R}_{~n}$ and after straightforward calculation, The master equation in Eq.~(14) of the main Letter takes the form
%\begin{equation}
%\begin{aligned}
%\mathcal{L}^X[\rho^t]=
%&\Gamma^+\left(\sigma^-_{\mathcal{S}_X}\rho^t\sigma^+_{\mathcal{S}_X}{-}\frac{1}{2}\lbrace \sigma^+_{\mathcal{S}_X}\sigma^-_{\mathcal{S}_X},\rho^t\rbrace\right)+\nonumber\\
%&\Gamma^-\left(\sigma^+_{\mathcal{S}_X}\rho^t\sigma^-_{\mathcal{S}_X}{-}\frac{1}{2}\lbrace \sigma^-_{\mathcal{S}_X}\sigma^+_{\mathcal{S}_X},\rho^t\rbrace\right)\\
%\mathcal{D}^{AB}[\rho^t]=
%&\Gamma^+\left(\sigma^-_{\mathcal{S}_A}\left[\rho^t,\sigma^+_{\mathcal{S}_B}\right]-\left[\rho^t,\sigma^-_{\mathcal{S}_B}\right]\sigma^+_{\mathcal{S}_A}\right)+\nonumber\\
%&\Gamma^-\left(\sigma^+_{\mathcal{S}_A}\left[\rho^t,\sigma^-_{\mathcal{S}_B}\right]-\left[\rho^t,\sigma^+_{\mathcal{S}_B}\right]\sigma^-_{\mathcal{S}_A}\right)
%\end{aligned}
%\end{equation}
\begin{equation}
\begin{aligned}
\dot{\rho}^t=&\mathcal{L}^A[\rho^t]+\mathcal{L}^B[\rho^t]+\mathcal{D}^{AB}[\rho^t]\\
\mathcal{L}^X[\rho^t]=
&\Gamma^+L[\sigma^-_{\mathcal{S}_X}](\rho^t)+\Gamma^-L[\sigma^+_{\mathcal{S}_X}](\rho^t)\\
\mathcal{D}^{AB}[\rho^t]=
&\Gamma^+\left(\sigma^-_{\mathcal{S}_A}\left[\rho^t,\sigma^+_{\mathcal{S}_B}\right]-\left[\rho^t,\sigma^-_{\mathcal{S}_B}\right]\sigma^+_{\mathcal{S}_A}\right)+\nonumber\\
&\Gamma^-\left(\sigma^+_{\mathcal{S}_A}\left[\rho^t,\sigma^-_{\mathcal{S}_B}\right]-\left[\rho^t,\sigma^+_{\mathcal{S}_B}\right]\sigma^-_{\mathcal{S}_A}\right)
\end{aligned}
\end{equation}
with $\rho^t$ the density matrix of  the overall system, $\gamma=g^2\tau$,  $\Gamma^\pm=2{\gamma}(1\pm\xi)$ and $L[\hat{o}](\rho){=} \hat{o}\rho\hat{o}^\dag {-}\frac{1}{2}\lbrace \hat{o}\hat{o}^\dag,\rho\rbrace$ for a generic operator $\hat{o}$. 
Needless to say, the reduced dynamics of ${\cal S}_A$ is not causally linked to that of $B$ due to the temporal ordering of the interactions between the sub-systems and a given sub-environment. The opposite is, quite obviously, not true. Explicitly
\begin{equation}
\label{MEAMEB}
\begin{aligned}
\dot{\rho}_{{\cal S}_A}^t&=\Gamma^{+}{L}[\sigma^-_{{\cal S}_A}]\rho_{{\cal S}_A}^t{+}\Gamma^-{L}[\sigma^+_{{\cal S}_A}]\rho_{{\cal S}_A}^t,\\
\dot{\rho}_{{\cal S}_B}^t&=\Gamma^{+}{L}[\sigma^-_{{\cal S}_B}]\rho_{{\cal S}_A}^t{+}\Gamma^-{L}[\sigma^+_{{\cal S}_B}]\rho_{{\cal S}_B}^t\\
&-i \frac{\Gamma^{+}{-} \Gamma^{-}}{2}\;\langle[\sigma^x_{{\cal S}_B},\sigma^y_{{\cal S}_A}\rho^t]-[\sigma^y_{{\cal S}_B},\sigma^x_{{\cal S}_A}\rho^t]\rangle_{{\cal S}_A},
\end{aligned}
\end{equation}
where the expectation value in the second line of Eq.~(\ref{MEAMEB}) is evaluated over the reduced state of sub-system ${\cal S}_A$. The expression for the heat flow from the overall system ${\cal S}$ can conveniently be given in terms of the elements $\rho^{ij}_{kl}=\langle i,j|\rho^t|k,l\rangle$ ($i,j,k,l=\uparrow,\downarrow$) of the two-particle density matrix. Explicitly
\begin{equation}
\dot Q(t)=4 \gamma \omega \left[ \xi(1 +  {\rho^{\uparrow\downarrow}_{\downarrow\uparrow}}+ {\rho^{\downarrow\uparrow}_{\uparrow\downarrow}})+ ( {\rho ^{\uparrow\uparrow}_{\uparrow\uparrow}}-   {\rho^{\downarrow\downarrow}_{\downarrow\downarrow}})\right]
\label{AdQ}\end{equation}
with $\{\ket{\uparrow},\ket{\downarrow}\}$ the eigenstates of $\hat \sigma^z$.
Eq. \eqref{AdQ} shows the direct dependence of the heat flow on the coherences of the $\mathcal S$ state. 
Similarly, 
\begin{eqnarray}
\dot Q_A(t){=}&&2 \gamma \omega\left[ 
(1{+}\xi)({\rho^{\uparrow\uparrow}_{\uparrow\uparrow}}{+}{\rho^{\uparrow\downarrow}_{\uparrow\downarrow}}){-}
(1{-}\xi) ( {\rho ^{\downarrow\uparrow}_{\downarrow\uparrow}}{+}{\rho^{\downarrow\downarrow}_{\downarrow\downarrow}})\right]\nonumber\\
\dot Q_B(t){=}&&2 \gamma \omega\left[ 
(1{+}\xi)({\rho^{\uparrow\uparrow}_{\uparrow\uparrow}}{+} {\rho ^{\downarrow\uparrow}_{\downarrow\uparrow}}){-}
(1{-}\xi) ({\rho^{\uparrow\downarrow}_{\uparrow\downarrow}}{+}{\rho^{\downarrow\downarrow}_{\downarrow\downarrow}})+\right.\nonumber\\
&&\left.2\xi(  {\rho^{\uparrow\downarrow}_{\downarrow\uparrow}}+ {\rho^{\downarrow\uparrow}_{\uparrow\downarrow}})\right]
\end{eqnarray}
Extending to a multipartite system $\mathcal{S}$ composed by $N$ qubits, the master equation becomes
\begin{equation}
\dot{\rho}^t=\sum_i^N \mathcal{L}^N[\rho^t]+\sum_{j>k}^N \mathcal{D}^{kj}[\rho^t].
\label{MEn}\end{equation} 
In Fig.(\ref{supp_1}) we report the heat flux and entropy variation behaviours for recycling environment in cases up to $5$ qubits. For $N=1$ this is the heat flow for a single qubit coupled to the environment with no recycling.
Looking at the heat flux (Fig.\ref{supp_1}{\bf (a)}), the initial growth is always greater in magnitude and lasts for longer as $N$ increases. This can be explained as a cooperative effect between the $N$ qubits which gets stronger as $N$ increases. Indeed, the master equation (\ref{MEn}) can be rearranged, writing a non-local coherent part due to a coupling mediated by the environment and collective decay channels with jump operators $J^\pm=\sum_k{\sigma^\pm_{\mathcal{S}_k}}$ \cite{zoller}. 
This effective interaction characterises the initial behaviour of the entropy variation and mutual information.  

%-----------------------------------------------------
\subsection{Example II: Indirect erasure via a qubit-harmonic oscillator system}\label{HO}
%----------------------------------------------------%-----------------------------------------------------
We now address the case of indirect erasure achieved through the use of a harmonic oscillator bridging the interaction of a qubit sub-system ${\cal S}_A$ and the multipartite environment ${\cal R}$. The interactions between the system qubit and the oscillator are described by resonant excitation-exchange interactions of the Jaynes-Cummings form. Specifically, we call $\hat b$ and $\hat b^\dag$ the bosonic annihilation and creation operators of the harmonic oscillator ${\cal S}_B$ and $\hat U_{AB}=e^{-i \hat H_{AB}\tau}$ the time-evolution operator generated by the Hamiltonian $\hat H_{AB}{=}\omega(\hat q^2{+}\hat p^2{+}\hat \sigma^z_{{\cal S}_A}/2){+}J(\hat q\hat \sigma^x_{{\cal S}_A}{+}\hat p\hat \sigma^y_{{\cal S}_A})$, with $q{=}(\hat b{+}\hat b^\dag)/\sqrt{2}$ and $p{=}i(\hat b^\dag{-}\hat b)/\sqrt{2}$ the position and momentum quadrature operators of the oscillator, respectively, 
and $\hat b(\hat b^\dag)$ the corresponding annihilation (creation) operator. 
 %$\hat H_{AB}=\omega(\hat b^\dagger \hat b + \omega{\hat \sigma^z_{{\cal S}_A}}/{2}+ J(\hat b\hat \sigma^+_{{\cal S}_A}+\hat b^\dag\hat \sigma^-_{{\cal S}_A})$. 
 This accounts for the intra-system collisions. Similarly, the interaction between the oscillator and the $n^{\rm th}$ element of the environment is described by the unitary operator $\hat U_{BR}=e^{-i \hat H_{BR}\tau}$ with $\hat H_{BR}$ given by $\hat H_{AB}$ with the replacements $J\to g$ and $\hat \sigma^{x,
y}_{{\cal S}_A}\to\hat \sigma^{x,y}_{{\cal R}_{~n}}$. By putting in place the procedure for deriving a continuous-time master equation, we find that the dynamics of system ${\cal S}$ is given by 
\begin{equation}
\label{eqHO}
\dot{\rho}^t{=}-i\left[\hat H_{AB},\rho^t\right]{+}\gamma_g\left(1{-}\xi\right)L[\hat b^\dag](\rho^t){+}\gamma_g\left(1{+}\xi\right)L[\hat b](\rho^t).%\left(\hat b^\dagger\rho^t\hat b-\frac{1}{2}\left\{aa^\dagger,\rho_{AB}\right\}\right)\nonumber\\
%&+\left(\frac{1+\xi}{2}\right)\left(a\rho_{AB}a^\dagger-\frac{1}{2}\left\{a^\dagger a,\rho_{AB}\right\}\right).
\end{equation}
The heat flow associated with the open-system nature of the evolution of ${\cal S}$ can then be expressed as functions of density matrix elements $\rho_{nm}^{kj}=\langle k,n|\rho^t|j,m\rangle$ ($j,k=\uparrow,\downarrow$ and $n,m\in{\mathbb Z}$) with $\{\ket{\uparrow},\ket{\downarrow}\}$ the eigenstates of $\hat\sigma^z_{{\cal S}_A}$ and $\{\ket{n}\}$ the Fock-state basis for ${\cal S}_B$. We have
\begin{eqnarray}\label{eqHOheat}
\dot Q(t)=
%\omega \gamma_g\left[\left(\frac{1-\xi}{2}\right)-\xi\sum_{n=0}^{\infty}n\left(\rho^{\uparrow\uparrow}_{nn}+\rho^{\downarrow\downarrow}_{nn}\right)\right].
%&&\omega \gamma_g\left[(1-\xi)\sum_{n=0}^{\infty}n\left(\rho^{\uparrow\uparrow}_{n-1n-1}+\rho^{\downarrow\downarrow}_{n-1n-1}\right)\right]+\\
%&&\omega \gamma_g\left[(1+\xi)\sum_{n=0}^{\infty}n\left(\rho^{\uparrow\uparrow}_{nn}+\rho^{\downarrow\downarrow}_{nn}\right)\right]
\omega \gamma_g\sum_{k=\uparrow,\downarrow}\sum_{n=0}^{\infty}n\left[(1{-}\xi)\rho^{kk}_{n{-}1n{-}1}+(1{+}\xi)\rho^{kk}_{nn}\right]
\end{eqnarray} 

%-----------------------------------------------------
\subsection{Heat flux and entropy variations versus coupling constants}\label{HO}
\begin{figure}[h!]
\begin{center}
%{\bf (a)}\hskip5.5cm{\bf (b)}\hskip5.5cm{\bf (c)}
\includegraphics[width=1.\columnwidth]{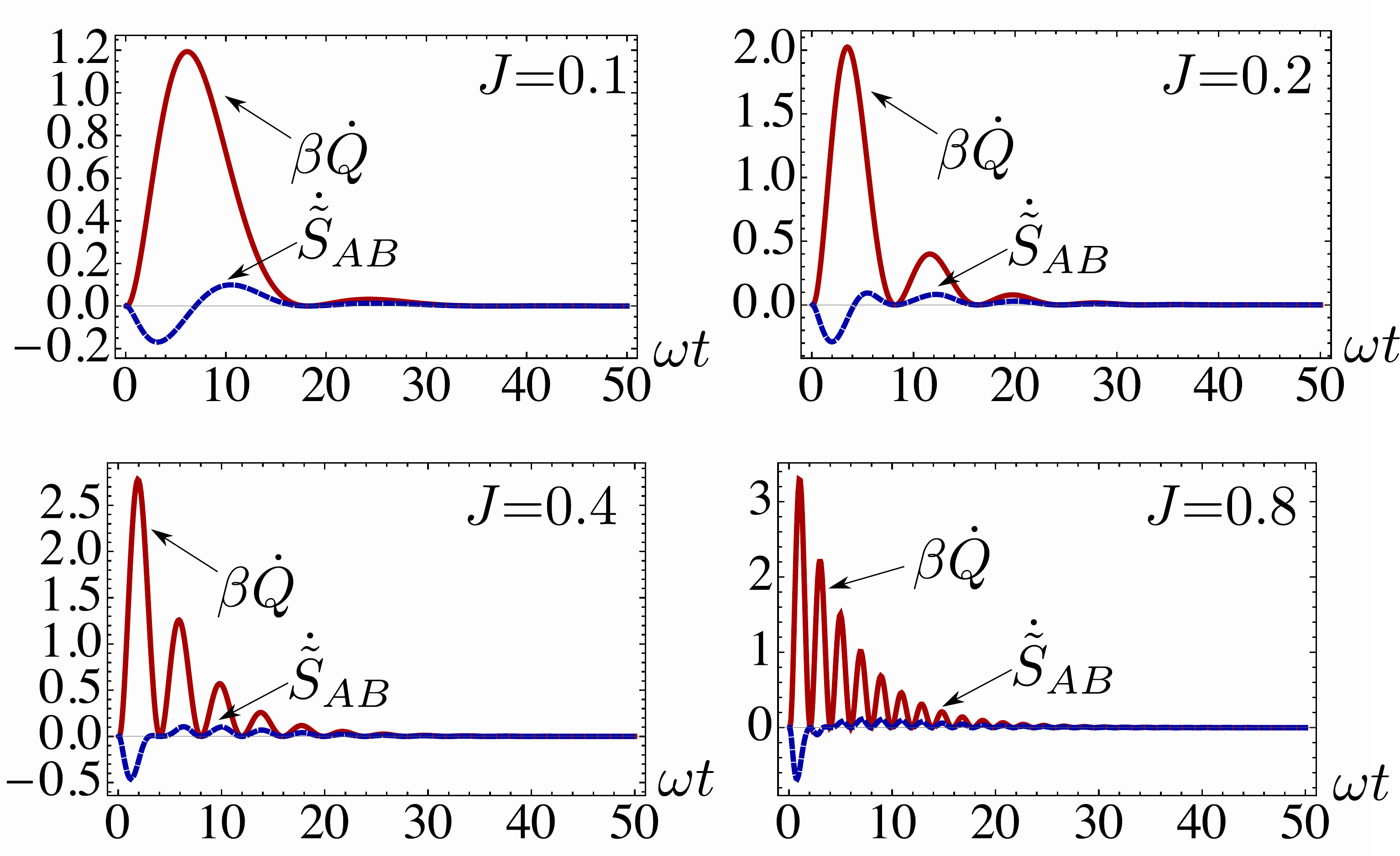}
\caption{(Color online) 
Total heat flow $\beta \dot{Q}$ (red) and the entropy change rate $\dot{S}$ (blue, dashed), for various values of coupling constant $J$ [in units of $\omega$] for the qubit-qubit configuration. We have set $\gamma_g/\omega{=}0.1$, taking $\mathcal{S}_B$ initially at the same temperature of the environment, $\beta=10$.}
\label{supp_2}
\end{center}
\end{figure}

%----------------------------------------------------%-----------------------------------------------------
Here we report the behavior of the heat flux and the entropy variations against the coupling rate $J$ between ${\cal S}_A$ and ${\cal S}_B$ in the indirect erasure protocol, for cusing on the qubit case for definiteness. Fig.~\ref{supp_2} shows the changes in the temporal behavior of both $\beta\dot{Q}$ and $\dot{\tilde S}_{AB}$ when $J=0.1,0.2,0.4$ and 0.8. When investigated within a fixed-size temporal window, the increase of the qubit-ancilla  coupling induces the appearance of an increasing number of oscillations both in the exchanged heat and the entropy changes. This is a clear result of the stronger coherent coupling between ${\cal S}_A$ and ${\cal S}_B$, which induces more substantial changes in the state of the former. 

\begin{figure}[h!]
\begin{center}
%{\bf (a)}\hskip5.5cm{\bf (b)}\hskip5.5cm{\bf (c)}
\includegraphics[width=1.\columnwidth]{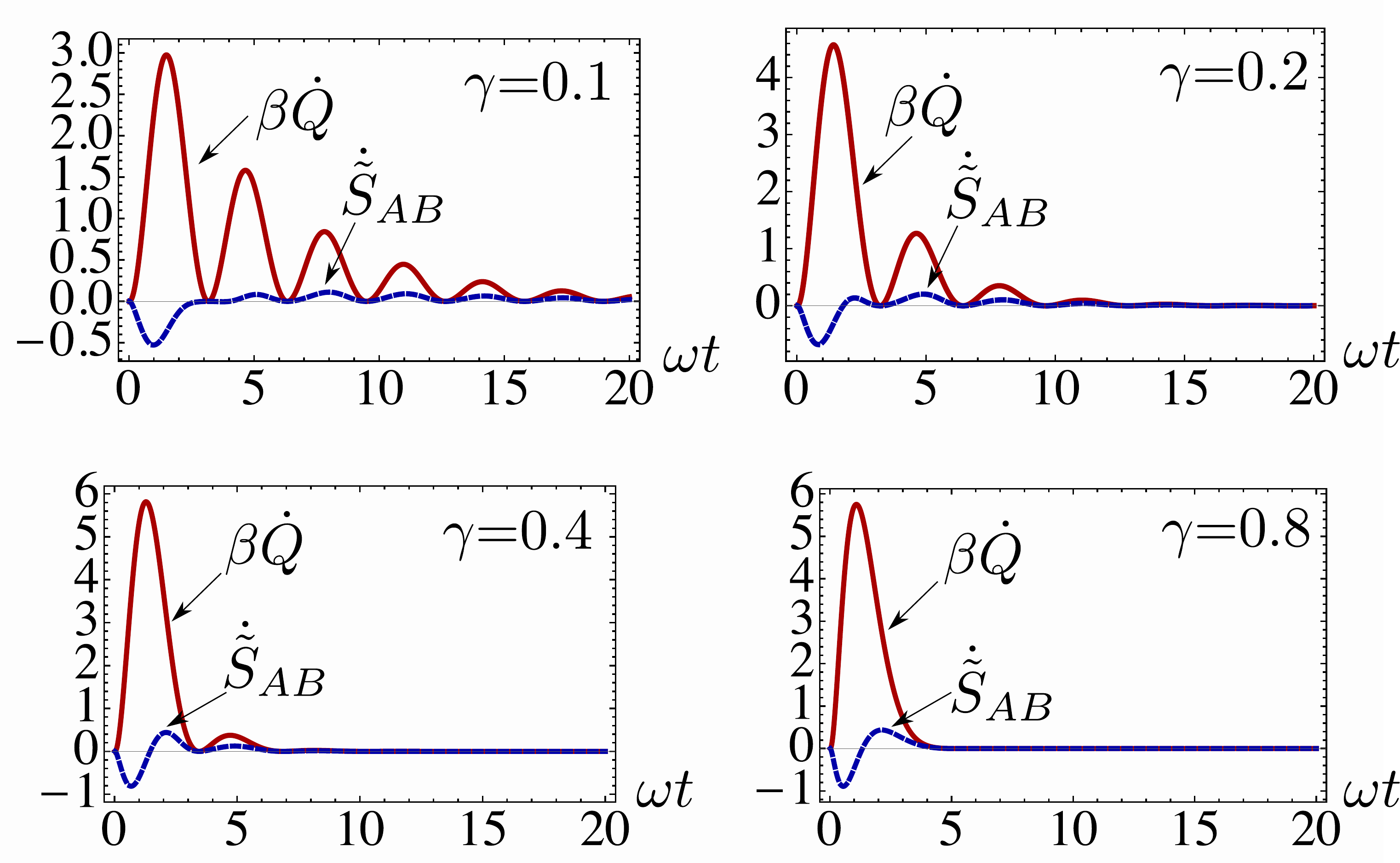}
\caption{(Color online) 
Total heat flow $\beta \dot{Q}$ (red) and the entropy change rate $\dot{S}$ (blue, dashed), for various values of rate $\gamma$ [in units of $\omega$] for a qubit-qubit configuration. We have set $J/\omega{=}0.5$ taking $\mathcal{S}_B$ initially at the same temperature of the environment, $\beta=10$.}
\label{supp_3}
\end{center}
\end{figure}

On the other hand, an increase in the effective damping rate $\gamma_g$ resulting from a ramped-up value of the ancilla-environment coupling rate $g$ results in a quicker equilibration of the state of ${\cal S}_A$: the system sees a fasted dissipative dynamics and adjusts to it very quickly. This is shown in Fig.~\ref{supp_3}.

%-----------------------------------------------------


\begin{thebibliography}{99}
\bibitem{Landauer}
R. Landauer, IBM J. Res. Dev. {\bf 5}, 183, (1961).
\bibitem{demone} C. Bennett, Int. J. Theor. Phys. {\bf 17}, 525 (1973); O. Penrose, {\it Foundations of statistical mechanics: A deductive treatment} (Pergamon, New Your, 1970).
\bibitem{experiments} A. B\'erut, A. Arakelyan, A. Petrosyan, S. Ciliberto, R. Dillenschneider, and E. Lutz, Nature {\bf 483}, 187 (2012); Y. Jun, M. Gavrilov, and J. Bechhoefer, Phys. Rev. Lett. {\bf 113}, 190601 (2014); A. O. Orlov, C. S. Lent, C. C. Thorpe, G. P. Boechler, and G. L. Snider, Jpn. J. Appl. Phys. {\bf 51}, 06FE10 (2012).
\bibitem{reviews} M. Esposito, U. Harbola, and S. Mukamel, Rev. Mod. Phys. {\bf 81}, 1665 (2009); M. Campisi, P. H\"anggi, and P. Talkner, {\it ibid.} {\bf 83}, 771 (2011).
\bibitem{progress} R. Dorner, S. R. Clark, L. Heaney, R. Fazio, J. Goold, and V. Vedral, Phys. Rev. Lett. {\bf 110}, 230601 (2013); L. Mazzola, G. De Chiara, and M. Paternostro, {\it ibid.} {\bf 110}, 230602 (2013); T. B. Batalh\~ao, et al., {\it ibid.} {\bf 113}, 140601 (2014); L. Mazzola, G. De Chiara, and M. Paternostro, Int. J. Quant. Inf. {\bf 12}, 1461007 (2014); F. Plastina, A. Alecce, T. J. G. Apollaro, G. Falcone, G. Francica, F. Galve, N. Lo Gullo, and R. Zambrini, Phys. Rev. Lett. {\bf 113}, 260601 (2014); M. Campisi, R. Blattmann, S. Kohler, D. Zueco, and P. H\"anggi, New. J. Phys. {\bf 15}, 105028 (2013); J. Goold, U. Poschinger, and K. Modi, Phys. Rev. E {\bf 90}, 020101R (2014); A. J. Roncaglia, F. Cerisola, and J. P. Paz, Phys. Rev. Lett. {\bf 113}, 250601 (2014); G. De Chiara, A. J. Roncaglia, and J. P. Paz, New J. Phys. {\bf 17}, 035004 (2015) .
\bibitem{recent} J. P. P. Silva, R. S. Sarthour, A. M. Souza, and I. S. Oliveira,  J. Goold, K. Modi, D. O. Soares-Pinto, and L. C. C\'eleri, arXiv:1412.6490 (2014).
\bibitem{Deffner} D. Kafri, and S. Deffner, Phys. Rev. A {\bf 86}, 044302 (2012); S. Deffner, and C. Jarzynski, Phys. Rev. X {\bf 3}, 041003 (2013); A. C. Barato, and U. Seifert, Phys. Rev. Lett. {\bf 112}, 090601 (2014); J. Parrondo, J. M. Horowitz, and T. Sagawa, Nat. Phys. {\bf 11}, 131 (2015). 
\bibitem{John} J. Goold, M. Huber, A. Riera, L. del Rio, and P. Skrzypczyk, arXiv:1505.07835 (2015).
\bibitem{reeb} D. Reeb and M. W. Wolf, New J. Phys. {\bf 16}, 103011 (2014).
\bibitem{goold} J. Goold, M. Paternostro, and K. Modi, Phys. Rev. Lett. {\bf 114}, 060602 (2015).
\bibitem{hilt} S. Hilt, S. Shabbir, J. Anders, and E. Lutz, Phys. Rev. E {\bf 83}, 030102 (2011).
\bibitem{CMs} J. Rau, Phys. Rev. {\bf 129}, 1880 (1963); R. Alicki and K. Lendi, {\it Quantum Dynamical
Semigroups and Applications}, Lecture Notes in Physics (Springer-Verlag, Berlin, 1987); M. Ziman {\it et al.}, Phys. Rev. A {\bf 65} , 042105 (2002); V. Scarani, M Ziman, P. Stelmachovic, N. Gisin and V. Bu\v{z}ek, Phys. Rev. Lett. {\bf88} , 97905 (2002); M. Ziman and V. Bu\v{z}ek, Phys. Rev. A {\bf 72}, 022110 (2005); M. Ziman, P. Stelmachovic, and V. Bu\v{z}ek, Open Sys. Inform. Dyn. {\bf 12}, 81 (2005).
G. Benenti, and G. M. Palma, Phys. Rev. A {\bf 75} 52110 (2007); G. Gennaro, G. Benenti and G. M. Palma, EPL {\bf 82} 2006, (2008), G. Gennaro, G. Benenti, and G. M. Palma, Phys. Rev. A {\bf 79}, 022105 (2009); G. Gennaro, S. Campbell, M. Paternostro, and G. M. Palma, {\it ibid.} {\bf 80}, 062315 (2009).
\bibitem{Palma_2012}
V. Giovannetti,  and G.M. Palma, Phys. Rev. Lett. {\bf 108}, 040401 (2012).
V. Giovannetti, G. M. Palma,  J. Phys. B {\bf 45}, 154003 (2012).
\bibitem{Ciccarello}
F. Ciccarello, G. M. Palma, and V. Giovannetti, Phys. Rev. A {\bf 87}, 040103(R) (2013); F. Ciccarello and V. Giovannetti, Phys. Scr. {\bf T153}, 014010 (2013).
\bibitem{McCloskey}
R. McCloskey and M. Paternostro, Phys. Rev. A {\bf 89}, 052120 (2014).
\bibitem{varieU} J. Spohn, J. Math. Phys. {\bf 19}, 1227 (1978); R. Alicki, J. Phys. A: Math. Gen. {\bf 12}, L103 (1979); W. Pusz, and S. L. Woronowicz, Commun. Math. Phys. {\bf 58}, 273 (1978). 
\bibitem{varieU2} S. Deffner, and E. Lutz, Phys. Rev. Lett. {\bf 107}, 140404 (2011); J. Anders, and V. Giovannetti, New J. Phys. {\bf 15}, 033022 (2013).
\bibitem{Breuer02} 
H.-P. Breuer and F. Petruccione, {\it The Theory of Open Quantum Systems} (Oxford University press, Oxford, 2002).
%\bibitem{Paternostro2014} J. Goold, M. Paternostro, and K. Modi, Phys. Rev. Lett. {\bf 114}, 060602 (2015).
%\bibitem{Esposito2006}
%M. Esposito and S. Mukamel, Phys. Rev. E {\bf 73}, 046129 (2006).
\bibitem{Rosario2013}
The relation is easily proved considering the form of the equilibrium state $\rho^{eq}$ and that, for a trace-preserving map, $\tr{\dot\rho}=0$. Indeed $\dot{S}(\rho|\rho^{eq})=\tr{\dot{\rho}\ln\rho-\dot\rho\ln\rho^{eq}}=\tr{\dot\rho\ln\rho}+\beta\tr{\dot\rho\,\hat H_{\cal S}}+\ln({\rm Tr}[{e^{-\beta\hat H_{\cal S}}}])\tr{\dot\rho}=-\dot{S}+\beta\tr{\dot\rho\,\hat H_{\cal S}}=-\dot{S}+\beta\dot E$. %C. Rodr\'iguez-Rosario, Th. Frauenheim, and A. Aspuru-Guzik, arXiv:1308.1245 (2013).
\bibitem{Vedral2002} V. Vedral, Rev. Mod. Phys. {\bf 74}, 197 (2002).
%\bibitem{Vedral_2010}
%B. Dakic, C. Brukner, and V. Vedral, Phys. Rev. Lett. {\bf 105}, 190502 (2010).
\bibitem{farace} S. Lorenzo, A. Farace, F. Ciccarello, G. M. Palma, and V. Giovannetti, Phys. Rev. A {\bf 91}, 022121 (2015).
\bibitem{nc} M. A. Nielsen and I. L. Chuang,  \textit{Quantum Computation and Quantum Information} (Cambridge University Press, Cambridge, U. K., 2000).
\bibitem{Herburt2004} F. Herbut,  J. Phys. A: Math. Gen. {\bf 37}, 3535 (2004).
\bibitem{zoller} K. Stannigel, P. Rabl, and P. Zoller, New J. Phys. {\bf 14}, 063014 (2012).

\end{thebibliography}
\end{document}